# Enhanced Third-Harmonic Generation in Diamond Photonic Crystal Slabs via Doubly Resonant Quasi-Bound States in the Continuum

Sangmin Ji[1, 2], Satoshi Iwamoto[2], and Hideki Gotoh[1]

[1]*Research Institute for Semiconductor Engineering, Hiroshima University, Higashihiroshima 739-8527, Japan*

[2]*Research Center for Advanced Science and Technology, The University of Tokyo, Tokyo 153-8904, Japan*

*smji@hiroshima-u.ac.jp*

## Abstract

We propose and numerically demonstrate doubly resonant third-harmonic generation (THG) in a diamond photonic crystal (PhC) slab, in which the fundamental harmonic (FH) and the third harmonic (TH) modes are simultaneously resonant within the same membrane. A hexagonal-lattice slab with triangular air holes is designed so that a K-point band-edge FH mode and a Γ-point quasi-bound-state-in-the-continuum (quasi-BIC) TH mode satisfy the frequency-tripling condition $3\omega_1 \approx \omega_3$. Modifying the hole shape from circular to equilateral triangular breaks the in-plane symmetry that otherwise forces the nonlinear coupling to vanish, thereby converting a TH mode with negligible overlap into one with finite while simultaneously reducing the required slab thickness. Guided by a closed-form expression for THG efficiency derived from coupled-mode theory, we design the unit cell and a PhC heterostructure cavity. Three-dimensional simulations of the designed cavity yield a normalized THG efficiency $\eta = 2.7\times10^{-7}$ W$^{-2}$ under moderate quality factors, which is projected to reach ~0.034 W$^{-2}$ at the fabrication-limited quality factor ($Q \approx$ 200,000). Because the operating wavelength is set by the lattice constant, this design, combined with the ultra-wide transparency window of diamond, can map a single geometry across various fabricable wavelengths, spanning from telecommunication bands to color-center-resonant visible and deep-UV outputs. These results establish a robust route toward efficient, monolithic on-chip frequency conversion in an all-diamond platform for quantum and nonlinear photonics.

**Highlights**

- Doubly resonant third-harmonic generation in a diamond photonic crystal cavity
- Band engineering for K-point band-edge mode and a Γ-point quasi-BIC to satisfy the frequency tripling condition
- Triangular air holes break the symmetry that forbids the nonlinear overlap
- A coupled-mode-theory figure of merit guides the cavity optimization
- A record-high THG conversion efficiency of 3.4 % demonstrated via 3D-FDTD simulation under fabrication-limited conditions

## 1. Introduction

With its ultra-wide transparency window extending below 220 nm and a relatively high refractive index (~2.4), diamond has emerged as an exceptional platform for on-chip photonics. Furthermore, its high thermal conductivity (22.9 W cm$^{-1}$ K$^{-1}$), large nonlinear refractive index ($n_2 = 1.3 \times 10^{-19}$ m$^2$ W$^{-1}$ [1]), and substantial Raman gain coefficient (~50 cm/GW [2]) make it highly suitable for nonlinear optics [3, 4]. Beyond these classical applications, diamond hosts various optically active color centers, most prominently the nitrogen-vacancy ($NV^-$) center, which exhibits long spin coherence times at room temperature and enable distributed quantum networking [5, 6]. However, a major spectral mismatch persists in these quantum architectures: spin–photon interactions primarily occur at visible wavelengths (e.g., 637 nm for $NV^-$ and 737 nm for $SiV^-$), whereas low-loss fiber-optic links operate in the telecommunication bands (1.3–1.55 µm). While heterogeneous schemes utilizing $\chi^{(2)}$ materials such as lithium niobate [7–9] have been widely deployed to bridge this gap, a monolithic, efficient wavelength conversion platform within diamond itself remains highly desirable for both quantum and classical photonic applications.

Since diamond is centrosymmetric, its lowest-order optical nonlinearity is governed by the third-order susceptibility ($\chi^{(3)}$). Because $\chi^{(3)}$ is intrinsically weak, efficient frequency conversion typically requires optical cavities to strongly enhance local fields, as demonstrated in stimulated Raman scattering [4, 10, 11], parametric oscillation [3], and four-wave mixing [12]. Among these processes, third-harmonic generation (THG) offers a direct, monolithic route for wide-range frequency tripling [13]. To maximize THG efficiency, while traditional schemes enhance only the fundamental harmonic (FH) resonance [14–21], the most effective strategy is to employ a doubly resonant cavity that simultaneously confines and couples both the FH and third harmonic (TH) modes. Although microring resonators can support such dual resonances [22], they demand stringent phase-matching conditions due to the nature of whispering-gallery modes. In contrast, photonic crystal (PhC) cavities alleviate these phase-matching constraints while providing wavelength-scale mode confinement. However, implementing a doubly resonant THG scheme in a PhC cavity is inherently challenging because the FH ($\omega$) and TH ($3\omega$) frequencies span far beyond any single photonic bandgap. At the TH frequency, the mode lies deep within the radiative light cone, preventing conventional bandgap confinement.

Bound states in the continuum (BICs) provide an effective solution to this bottleneck; being embedded within the radiative continuum yet decoupled from it, BICs support high-$Q$ resonances outside the photonic bandgap across a wide tunable frequency range [23]. While BIC-based PhC metasurfaces have recently attracted considerable attention for harmonic generation [14–16], they generally assume plane-wave illumination. Since high-intensity, localized pumping is essential for efficient THG, spatially confined PhC cavities offer a distinct advantage. Although doubly resonant second-harmonic generation using BICs in PhC heterostructure cavities has been demonstrated in non-centrosymmetric III–V materials [24, 25], a doubly resonant THG scheme, however, has not to our knowledge been established for a centrosymmetric material such as diamond, where the symmetry constraints are more demanding: the two resonances are separated by a factor of three rather than two, and the $\chi^{(3)}$ tensor imposes different selection rules from those of $\chi^{(2)}$.

In this work, we propose and numerically demonstrate a doubly resonant hexagonal-lattice diamond PhC heterostructure cavity that simultaneously enhances the FH and TH modes. We overcome the challenge of vanishing nonlinear coupling by performing band engineering that tailors the nonlinear overlap between FH and TH modes via the $\chi^{(3)}$ tensor. To guide the cavity optimization, we derive a closed-form expression for the THG efficiency of a doubly resonant cavity and use it as a figure of merit to optimize the PhC unit cell. By embedding the optimized cell into a heterostructure cavity, we numerically demonstrate a doubly resonant THG efficiency that substantially outperforms that of previously reported structures. Because our design strategy relies strictly on lattice symmetry and a material-independent FOM, it provides a general and scalable recipe for enhancing $\chi^{(3)}$ interactions across a broad range of integrated photonic platforms and wavelengths.

## 2. Doubly Resonant THG Efficiency from Coupled-Mode Theory

First, we formulate the THG efficiency of a doubly resonant cavity by temporal coupled-mode theory (CMT). Although CMT frameworks for doubly resonant cavities have been developed in prior work [22, 26, 27], the precise relationship between the cavity's generalized parameters and the THG efficiency remains underexplored. We model the system as two coupled cavity modes: the FH mode of amplitude $a_1$ (frequency $\omega_1$, total lifetime $\tau_1$, external coupling $\tau_{ex,1}$) driven by an input $s_+$, and the TH mode of amplitude $a_3$ (frequency $\omega_3$, lifetime $\tau_3$, coupling $\tau_{ex,3}$). Normalizing $|a_k|^2$ to the stored energy and $|s_{k+(-)}|^2$ to the incoming (outgoing) power, temporal CMT gives

$$\frac{da_1}{dt} = \left[i\omega_1(1 - \alpha_{11}|a_1|^2 - \alpha_{13}|a_3|^2) - \frac{1}{\tau_1}\right] a_1 - i\omega_1\beta_1\,(a_1^*)^2 a_3 + \sqrt{\left(\frac{2}{\tau_{ex,1}}\right)}\, s_+ \tag{1}$$

$$\frac{da_3}{dt} = \left[i\omega_3(1 - \alpha_{33}|a_3|^2 - \alpha_{31}|a_1|^2) - \frac{1}{\tau_3}\right] a_3 - i\omega_3\beta_3 a_1^3 + \sqrt{\left(\frac{2}{\tau_{ex,3}}\right)}\, s_{3+} \tag{2}$$

$$s_- = -s_+ + \sqrt{\frac{2}{\tau_{ex,1}}}\, a_1, \quad s_{3-} = \sqrt{\frac{2}{\tau_{ex,3}}}\, a_3 \tag{3, 4}$$

, where $\beta_k$ is the nonlinear coupling coefficient. In the weak-conversion regime, we neglect the self- and cross-phase-modulation terms $\alpha_{ab}$, drop the back-conversion term $\beta_1$ in Eq. (1), and set the TH drive $s_{3+} = 0$ in Eq. (2). Under the harmonic approximation $s_+(t) = s_+(\omega)e^{i\omega t}$, $a_1 = a_1(\omega)e^{i\omega t}$, $a_3 = a_3(3\omega)e^{3i\omega t}$, the steady state becomes

$$\left[i(\omega-\omega_1)+\frac{1}{\tau_1}\right]a_1 = \sqrt{\frac{2}{\tau_{ex,1}}}\,s_+,\quad \left[i(3\omega-\omega_3)+\frac{1}{\tau_3}\right]a_3 = -i\omega_3\beta_3\,a_1^3. \tag{5, 6}$$

Defining the normalized THG efficiency $\eta$ [22, 26] $\equiv P_{3-}(3\omega) / [P_{1+}(\omega)]^3 = |s_{3-}|^2 / |s_+|^6$ and eliminating $a_1$ and $a_3$ yields

$$\eta = \frac{16\omega_3^2|\beta_3|^2\tau_1^6\tau_3^2}{\tau_{ex,1}^3\tau_{ex,3}}\cdot L_1(\omega)^3L_3(\omega) \tag{7}$$

, where $L_1(\omega) = \left[1+\left(\tau_1(\omega-\omega_1)\right)^2\right]^{-1}$ and $L_3(\omega) = \left[1+\left(\tau_3(3\omega-\omega_3)\right)^2\right]^{-1}$ are the FH and TH Lorentzian line shapes. Introducing the loaded quality factors $Q_1 = \omega_1\tau_1/2$ and $Q_3 = \omega_3\tau_3/2$ and assuming negligible absorption so that each mode is critically coupled ($\tau_a = \tau_{ex,a}$), Eq. (7) reduces to the compact form

$$\eta = 256\left(\frac{\omega_3}{\omega_1^3}\right)Q_1^3\,Q_3|\beta_3|^2L_1(\omega)^3L_3(\omega). \tag{8}$$

Under the doubly resonant condition ($\omega = \omega_1$ and $3\omega = \omega_3$), the line shapes equal unity, so $\eta \propto Q_1^3Q_3|\beta_3|^2$. Equation (8) is the figure of merit (FOM) that guides the optimization below. Here, the nonlinear overlap is defined as [26],

$$\beta_3 = \frac{1}{8}\frac{\int[\boldsymbol{P}^{(3)}(3\omega)]\cdot\boldsymbol{E}^*(3\omega)dV}{[\int\varepsilon(\omega)|\boldsymbol{E}(\omega)|^2dV]^{\frac{3}{2}}\,[\int\varepsilon(3\omega)|\boldsymbol{E}(3\omega)|^2dV]^{\frac{1}{2}}} \tag{9}$$

where the nonlinear polarization vector $\boldsymbol{P}^{(3)}(3\omega)$ is

$$P_i^{(3)}(3\omega) = \varepsilon_0\Sigma_{jkl}\chi_{ijkl}^{(3)}E_j(\omega)E_k(\omega)E_l(\omega). \tag{10}$$

A nonzero $\beta_3$ thus requires spatial symmetry compatibility between the nonlinear polarization $\boldsymbol{P}^{(3)}(3\omega)$ driven by the FH field and the TH mode field, motivating the symmetry-guided design of Section 3.

## 3. Design of PhC for Doubly Resonant THG

As a base design for the doubly resonant PhC, we consider a hexagonal-lattice PhC slab. The nonlinear overlap $\beta_3$ is the spatial overlap between the nonlinear polarization $\boldsymbol{P}^{(3)}(3\omega)$, formed from the FH mode field $\boldsymbol{E}(\omega)$, and the TH mode field $\boldsymbol{E}^*(3\omega)$; a nonzero $\beta_3$ therefore requires the symmetry of $\boldsymbol{P}^{(3)}(3\omega)$ to be compatible with that of the TH mode. We adopt the TE polarization ($E_x$, $E_y$, $H_z$) and use the lowest band-edge mode at the K point as the FH mode ($\omega_1$); this mode is readily confined in a PhC heterostructure cavity owing to the relatively large TE bandgap of the hexagonal lattice [24]. The TE bandgap opens between the K point of the first (dielectric) band and the M point of the second (air) band. Compared with the M-point mode, the K-point band-edge mode is preferable for the nonlinear interaction: its electric field is more strongly concentrated in the dielectric region, and its lower frequency keeps the TH frequency below the onset of higher-order diffraction, which would otherwise complicate the band engineering. Because it is a band-edge mode, the FH mode can attain an arbitrarily large quality factor $Q_1$, which is critical for a large conversion efficiency $\eta$ [Eq. (8)].

As candidates for the TH mode ($\omega_3$), we first identify nondegenerate modes at the Γ point, which are symmetry-protected BICs because their mode symmetry is incompatible with the radiation continuum [23, 28]. We restrict attention to modes sharing the same sign of band curvature ($d^2\omega/dk^2 < 0$) as the FH mode, to ensure that both band maxima fall within the same heterostructure mode gap and are thus co-localized; modes of opposite curvature could not be confined by a common mode gap. Figure 1(a) shows the TE band structure of the circular-hole hexagonal lattice (refractive index $n = 2.42$, hole radius $r = 0.27a$, lattice constant $a$), calculated by the two-dimensional plane-wave expansion method. There are two such bands satisfying the above requirements in the range of interest, $\omega_3a/(2\pi c) = 0.75$–$1.2$, which can be spectrally aligned to satisfy $\omega_3 = 3\omega_1$ where $\omega_1a/(2\pi c)$ typically lies in 0.25–0.4. We label these two modes as mode I and mode II. As depicted in Fig. 1(d), mode I meets the symmetry condition for a nonzero $\beta_3$ with the FH mode. However, in a three-dimensional slab, it requires an impractically large thickness, $t = 1.6a$, for $\omega_3 \approx 3\omega_1$. On the other hand, mode II possesses zero $\beta_3$ with the FH mode in the circular-hole lattice owing to its mode symmetry. It should be noted that $\beta_3$ for mode II in three-dimensional case could be nonzero but still very small. We therefore introduce asymmetry by continuously deforming the circular hole into an equilateral triangular hole while keeping the hole area constant, as depicted in Fig. 1(b). In the triangular-hole PhC, Mode II transforms into Mode II$^{(t)}$, which acquires a nonzero $\beta_3$ with the FH mode owing to the broken symmetry (Fig. 1(d)). In addition, the triangular hole lowers the frequency of Mode II as shown in Fig. 1(c), shifting it into the range suitable for THG.

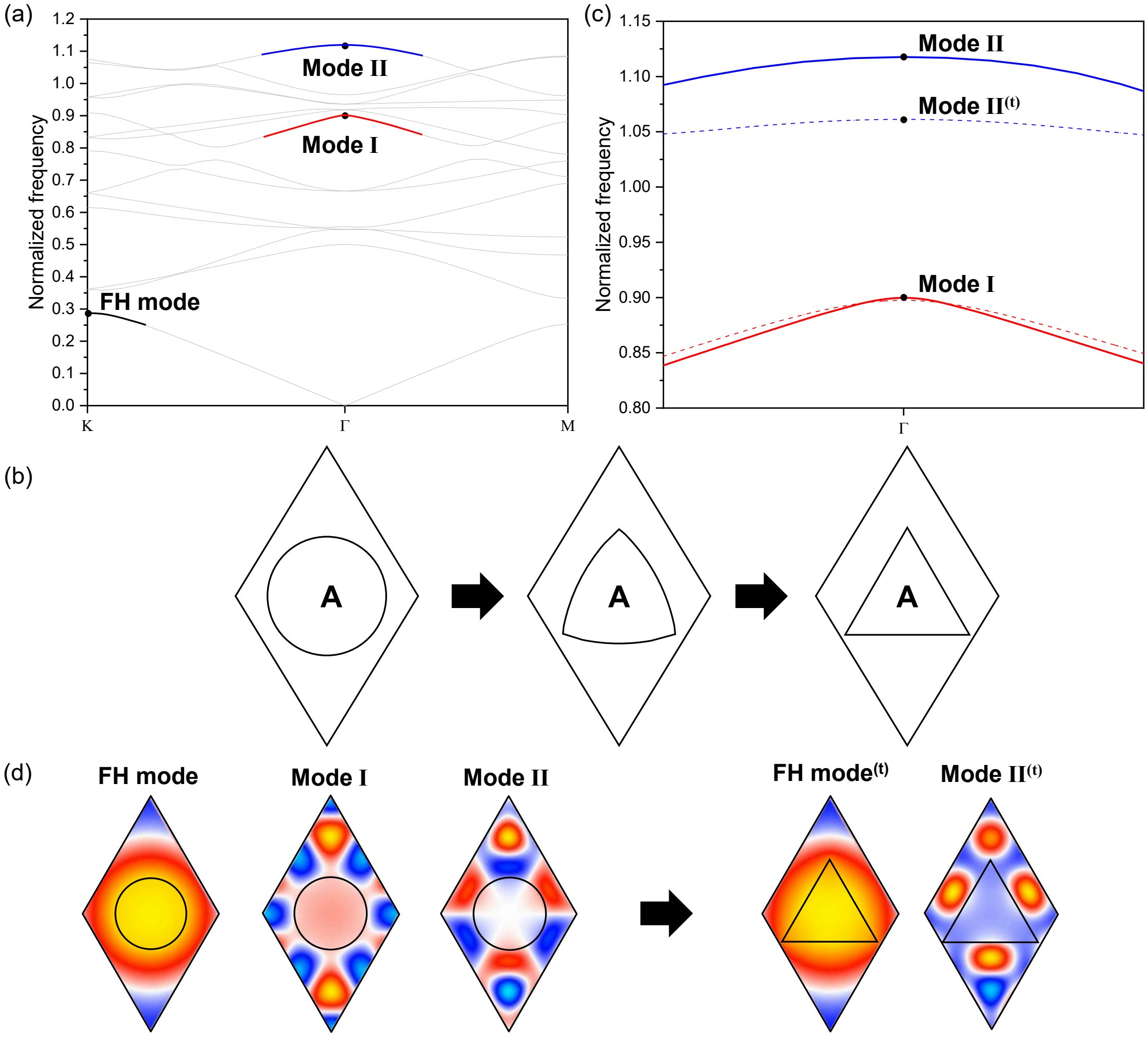

Fig. 1. (a) 2D TE band structure of the circular-hole hexagonal lattice (n = 2.42, $r = 0.27a$). The FH mode and two BICs (Mode I and Mode II) are indicated. (b) Continuous deformation of a circular hole into an equilateral triangular hole, with its centroid fixed at the unit-cell center, at constant area $A = \pi r^2$. (c) Frequency shift of Mode I and Mode II upon the circular-to-triangular-hole transformation. Solid (dashed) lines indicate the bands for circular (triangular) hole. (d) Spatial distribution ($H$ field perpendicular to the PhC) of the FH mode, Mode I, and mode II in the circular-hole lattice (left), and of FH mode$^{(t)}$ and Mode II$^{(t)}$ in the triangular-hole lattice (right).

Next, guided by the FOM of Eq. (8), we maximize $Q_1^3 Q_3 |\beta_3|^2$ over the three-dimensional unit-cell geometry. We adopt the material dispersion of single-crystal diamond, n = 2.38 (2.42) for the FH (TH) modes near 1550 nm (517 nm) [13], and set the third-order susceptibility tensor to $\chi_{1111} = 3\chi_{1122} = 2.1 \times 10^{-21}$ m$^2$V$^{-2}$ [1, 4], where the relation $\chi_{1111}$ $= 3\chi_{1122}$ is the isotropic approximation for the cubic crystal. Under this approximation, the THG polarization reduces to $\boldsymbol{P}^{(3)}(3\omega) = \varepsilon_0\chi_{1111}(\boldsymbol{E}\cdot\boldsymbol{E})\boldsymbol{E}(\omega)$ [1, 15]. Since $\boldsymbol{P}^{(3)}(3\omega)$ is then strictly collinear with $\boldsymbol{E}(\omega)$, any cross-polarized nonlinear coupling between the TE-FH mode and a TH mode of different polarization vanishes. Coupling to TM-like TH modes is further suppressed by the slab mirror symmetry, which enforces $P_z^{(3)} = 0$ for in-plane TE fields in a cubic crystal. The nonlinear overlap $\beta_3$ is therefore determined by the co-polarized TE–TE coupling. As a band-edge mode at the K point below the light line, the practical limit on $Q_1$ is $Q_{\text{bound}}$, determined by fabrication imperfections and material absorption. The TH mode is a Γ-point BIC located deep within the light cone, and $Q_3$ accordingly depends on the structure parameters due to the possible band mixing. We therefore adopt $Q_3|\beta_3|^2$ as the unit-cell FOM, and include material absorption with $Q_{\text{abs}} = 2 \times 10^5$ [29] as an upper bound for both $Q_1$ and $Q_3$. Figure 2 shows the dependence of FOM $Q_3|\beta_3|^2$ on the triangular-hole size $d$ for various thicknesses $t$. The FOM is maximized at $(d/\sqrt{3})/a = 0.4375$, 0.455, 0.4625, for $t/a = 0.50, 0.55, 0.60$, respectively. A further increase in $d$ makes the TH mode leaky, causing a rapid drop in the FOM. Note that the hole size $d$ that maximizes the FOM increases with $t$, while $d$ satisfying the tripling condition $\omega_3 = 3\omega_1$ decreases with $t$. Combining these opposite trends, $t/a = 0.55$ yields the highest FOM among the thicknesses considered. The corresponding three-dimensional band diagram for the designed structure $(t/a, d/a) = (0.55, 0.451\sqrt{3})$ is shown in Fig. 3.

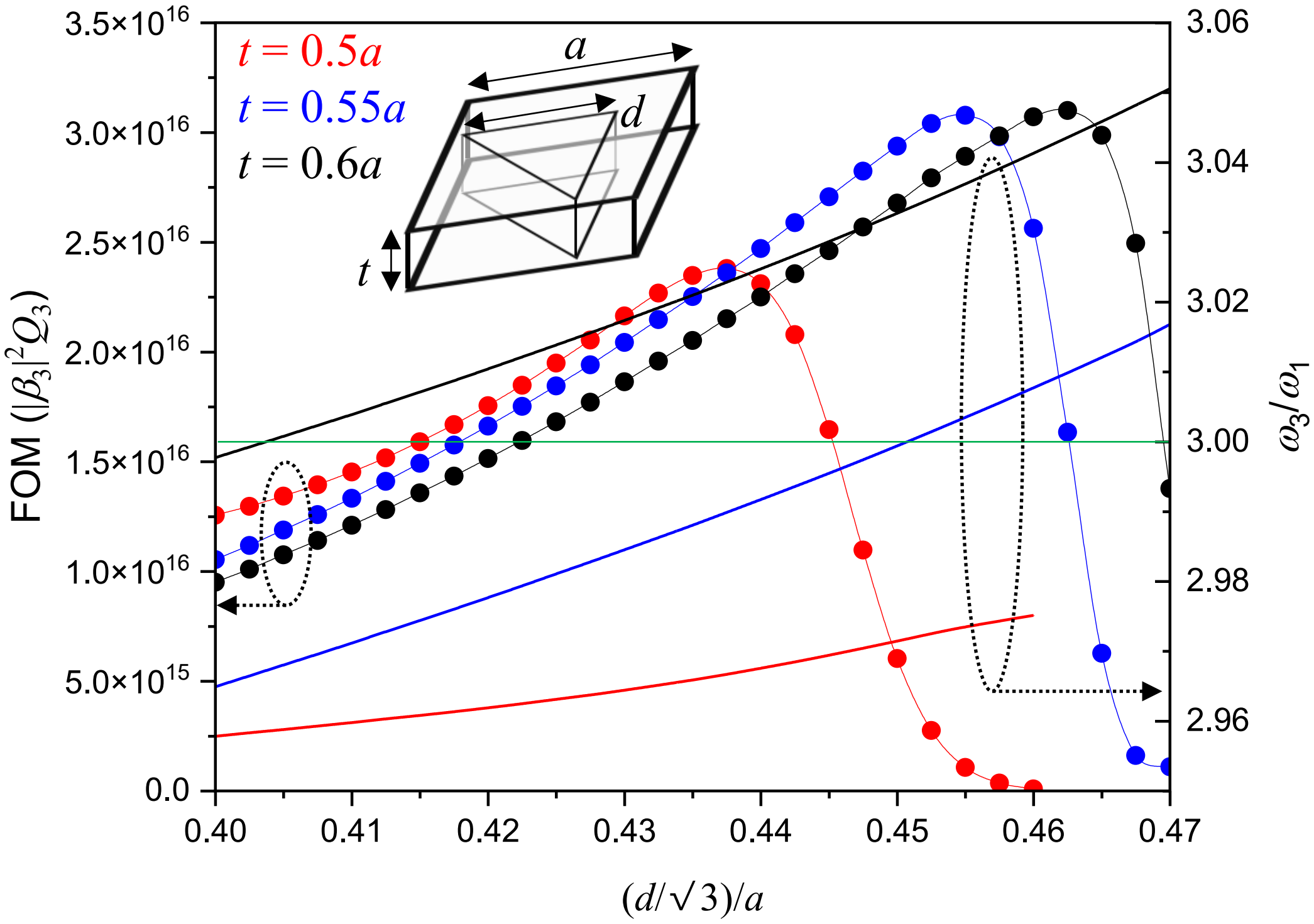

**Fig. 2.** Figure of merit $Q_3|\beta_3|^2$ (left axis) and the TH-to-FH frequency ratio $\omega_3/\omega_1$ (right axis) as a function of hole size $(d/\sqrt{3})/a$ for slab thickness $t/a = 0.5$, 0.55, and 0.6 (Inset: schematic of the unit cell structure with lattice constant $a$, hole size $d$, and slab thickness $t$). The frequency tripling condition $\omega_3/\omega_1 = 3$ is indicated as a green line.

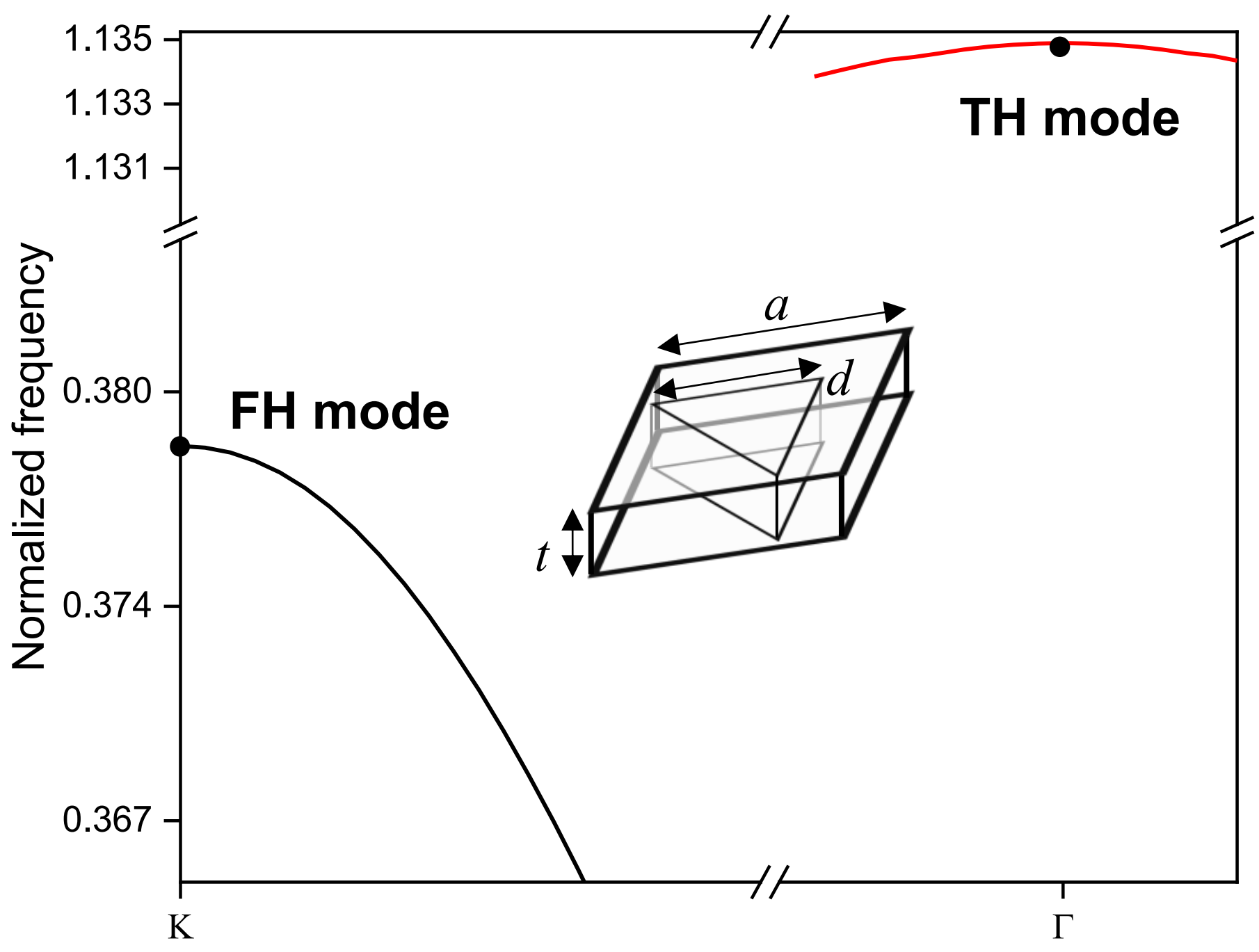

**Fig. 3.** Three-dimensional band structure of the triangular-hole diamond PhC slab ($d = 0.451\sqrt{3}a$, $t = 0.55a$). The equivalent filling factor for the circular-hole PhC is $r = 0.29a$.

## 4. Design of Doubly Resonant PhC Heterostructure Cavity for Enhancing THG

Based on the PhC cell design in the previous section, we performed three-dimensional finite difference time domain (3D-FDTD) simulations to verify the design in realistic PhC heterostructure cavities. We used the wavelength-dependent refractive index of diamond [13] in the simulations. Figure 4 shows a schematic of the designed PhC heterostructure cavity comprising core, transition, and cladding layers. These layers contain $n_{core}$, $n_{trans}$, and $n_{clad}$ periods, respectively. The slab thickness and the triangular hole sizes in the core, transition, and cladding regions are set to ($t$, $d_{core}$, $d_{trans}$, $d_{clad}$) = ($0.55a$, $0.46\sqrt{3}a$, $0.44\sqrt{3}a$, $0.4\sqrt{3}a$), where $a$ is the period of hexagonal PhC array. The photonic bands of the cladding region are pulled down to the lower frequencies to position the band maxima of the core region within the mode gap, enabling the simultaneous localization of both FH and TH modes in the core region. The role of the transition layer is to smooth the field distributions in real space, thereby sharpening the momentum-space spectrum, improving the cavity mode quality factor [30]. We calculated the mode properties, namely eigenfrequencies, quality factors, and field distributions of FH and TH modes, and evaluated the nonlinear field overlap $|\beta_3|^2$ by Eq. (9).

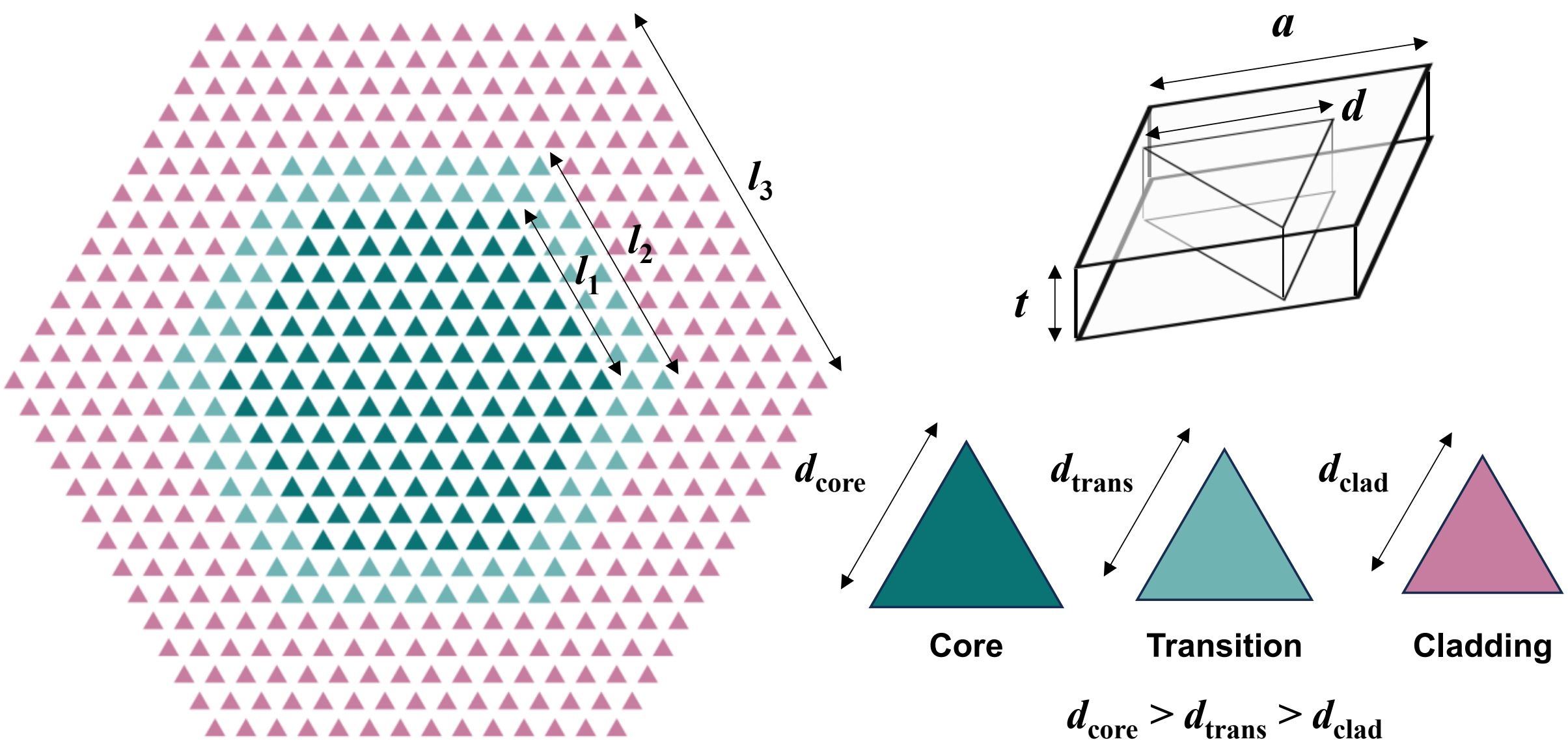

**Fig. 4.** Schematic of the PhC heterostructure cavity comprising core, transition and cladding layers. Each layer contains $n_{core}$, $n_{trans}$, and $n_{clad}$ periods respectively, such that the hexagon side lengths are $l_1 = n_{core}a$, $l_2 = (n_{core} + n_{trans})a$, and $l_3 = (n_{core} + n_{trans} + n_{clad})a$. The side length of equilateral triangular holes for each layer are $d_{core}$, $d_{trans}$, and $d_{clad}$, respectively, where the graded hole size $d_{core} > d_{trans} > d_{clad}$ shifts the band maxima of core region into the mode gap of the cladding region.

As shown in Fig. 5(a), the quality factor of FH mode $Q_1$ increases exponentially with the number of cladding layers $n_{clad}$. This is true for every combination of ($n_{core}$, $n_{trans}$, $n_{clad}$) because the FH mode is a K-point band-edge mode that falls within the heterostructure mode gap [24, 30]. We also found that, for the structures considered, the optimum $n_{trans} = 2$ through independent simulations; we therefore fixed ($n_{trans}$, $n_{clad}$) = (2, 5) in the following simulations to reduce the computational cost; $n_{clad} = 5$ provides a representative $Q_1 \approx 4000$ at a tractable cost, with the extrapolation to larger $n_{clad}$ given by Fig. 5(a). We then evaluated the reduced FOM $Q_3|\beta_3|^2$ by varying $n_{core}$. While $Q_3$ increases with $n_{core}$ due to the quasi-BIC nature [24, 30, 31] of the mode (Fig. 5(b)), the nonlinear overlap $|\beta_3|^2$ peaks near $n_{core}$ = 3–4 and decreases for $n_{core} > 4$ (Fig. 5(c)), because for small $n_{core}$ the TH mode is not yet fully developed as a BIC, so $|\beta_3|^2$ first rises with $n_{core}$, whereas for larger $n_{core}$ the growing mode volume dilutes the normalized field overlap so that $Q_3|\beta_3|^2$ is maximized at $n_{core} = 6$ (Fig. 5(d)).

In Fig. 6, we present the 3D-FDTD result for the designed PhC heterostructure ($a$ = 0.584 µm, $t = 0.55a$, ($d_{core}$, $d_{trans}$, $d_{clad}$) = ($0.46\sqrt{3}a$, $0.44\sqrt{3}a$, $0.4\sqrt{3}a$), ($n_{core}$, $n_{trans}$, $n_{clad}$) = (6, 2, 5)). Both harmonics are co-confined within the hexagonal core: the FH mode ($\lambda_1$= 1.541 µm, $f_1$ = 194.6 THz, $Q_1 \approx 4000$), and the TH mode ($\lambda_3$ = 514 nm, $f_3$ = 583.8 THz, $Q_3 \approx 1622$). The frequency ratio is $f_3/f_1 = 2.9996$, representing a deviation of 0.01% from exact frequency tripling, well within the cavity linewidth of the TH mode. The calculated nonlinear overlap reaches $|\beta_3|^2 = 5.8 \times 10^6$ J$^{-2}$. Inserting these values into Eq. (8) gives a normalized THG efficiency $\eta = 2.7 \times 10^{-7}$ W$^{-2}$. Note that this value of $\eta$ corresponds to $Q_1$ = 4000, obtained with $n_{clad}$ = 5. Considering the simulated trend shown in Fig. 5(a), $Q_1$ could reach the fabrication-limited value with $n_{clad} > 13$, i.e., $Q_1 \sim Q_{bound} = 2.0 \times 10^5$. Since $\eta$ scales as $Q_1^3$, and $Q_3$ and $\beta_3$ saturate with increasing $n_{clad}$ (confirmed by independent simulations), $\eta$ could be as large as ~0.034 W$^{-2}$.

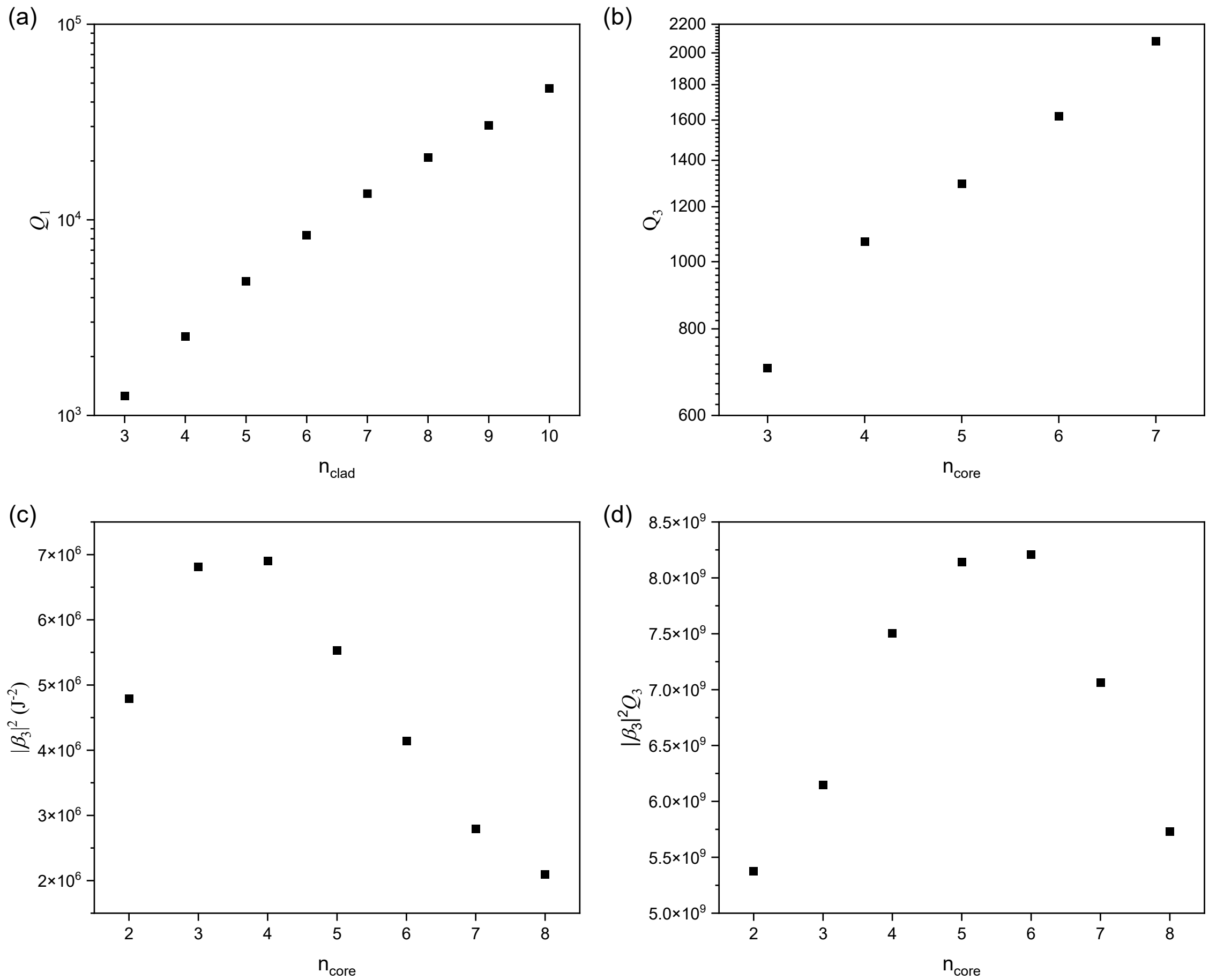

**Fig. 5.** (a) FH mode quality factor $Q_1$ as a function of $n_{clad}$. (b) TH mode quality factor $Q_3$ as a function of $n_{core}$. (c) Nonlinear overlap $|\beta_3|^2$ and (d) reduced figure of merit $Q_3|\beta_3|^2$ as a function of $n_{core}$. Despite the monotonic increase in $Q_3$ for larger $n_{core}$, the drop of $|\beta_3|^2$ limits the optimal $n_{core}$ to 6.

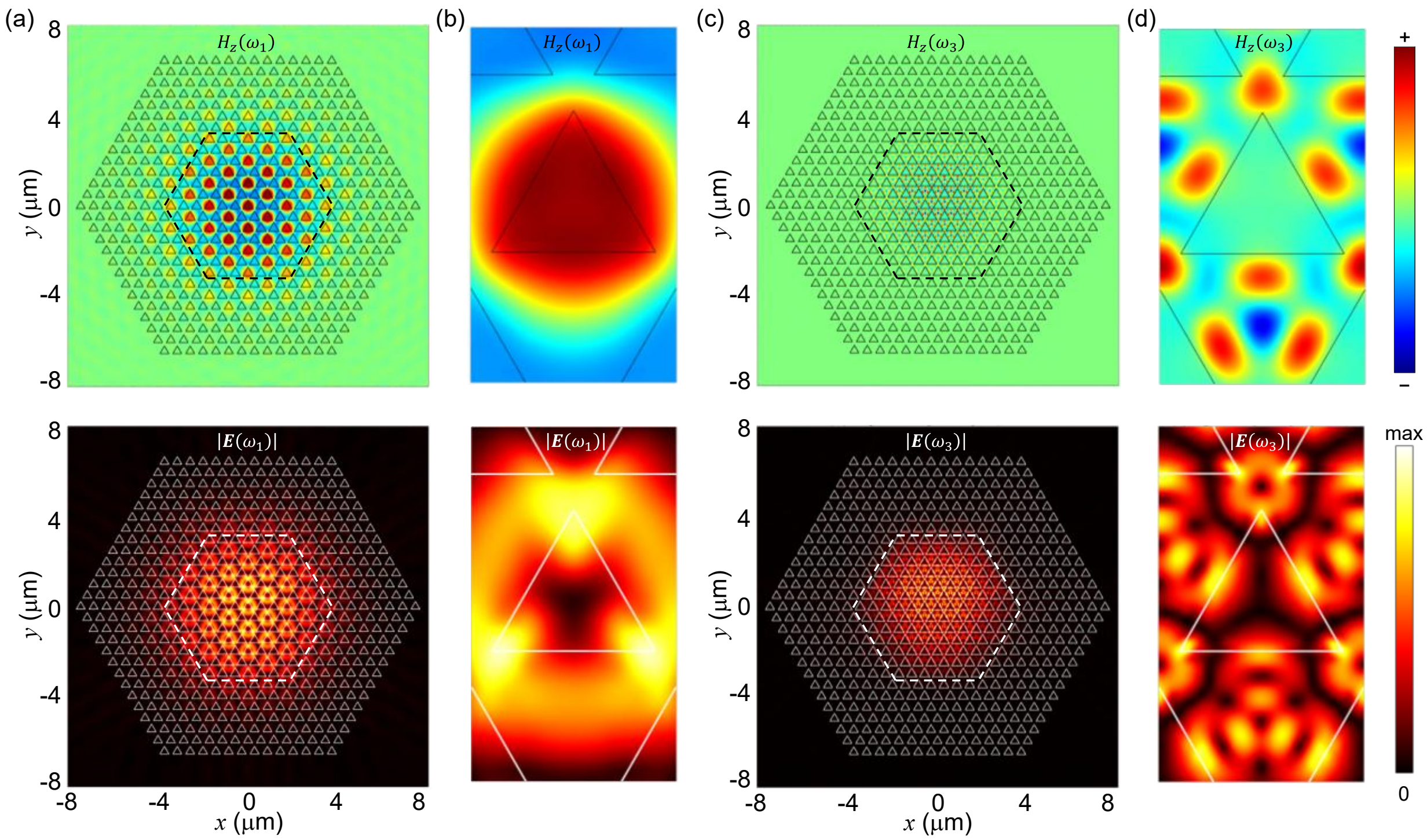

**Fig. 6.** 3D-FDTD simulated PhC heterostructure cavity ($a$ = 0.584 µm, $t$ = 0.55$a$, ($d_{\mathrm{core}}$, $d_{\mathrm{trans}}$, $d_{\mathrm{clad}}$) = (0.46$\sqrt{3}a$, 0.44$\sqrt{3}a$, 0.4$\sqrt{3}a$), ($n_{\mathrm{core}}$, $n_{\mathrm{trans}}$, $n_{\mathrm{clad}}$) = (6, 2, 5)). Top row: $H_z$ Field distributions; bottom row: $|\boldsymbol{E}|$ field distributions, for the FH mode[(a), (b)] and the TH mode [(c), (d)], shown as full cavity view and zoomed core view, respectively. The core region boundary is indicated by a dashed line.

## 5. Discussion and Benchmarking of the Results

To benchmark the performance of our PhC heterostructure cavity against related works, we assume resonant pumping focused on the core region (45 µm$^2$) at the FH frequency. For simplicity, we consider pumping power $P_{\mathrm{pump}}$ = 1 W, namely 2.2 MW/cm$^2$. The THG conversion efficiency $\eta_{\mathrm{conv}} = P_{\mathrm{THG}}/P_{\mathrm{pump}}$, evaluated at a specific pumping intensity $I_{\mathrm{pump}} = P_{\mathrm{pump}} / S$ (where $S$ is the device area), is commonly used as a performance benchmark in the literature. In particular, theoretical works on metasurface-based THG [14, 15] normally assume plane-wave incidence on periodic cells and therefore define $I_{\mathrm{pump}}$ as $P_{\mathrm{pump}}$ at the input port divided by the unit cell area. In table 1, we summarize representative $\eta_{\mathrm{conv}}$ from related works. Note that most THG works employ silicon as a nonlinear material, which has $\chi^{(3)} = 2.45 \times 10^{-19}$ m$^2$ V$^{-2}$ [14–16], two orders of magnitude larger than that of diamond. Because the conversion efficiency roughly scales as $(\chi^{(3)})^2$ (see the definition in Eq. (9)), we also compare $\eta_{\mathrm{conv}}$ divided by $P_{\mathrm{pump}}\cdot(\chi^{(3)})^2$, to account for differences in material and pumping power.

Even at conservative quality factor $Q_1$ of 4000, our proposed design outperforms most previously reported structures. For a fabrication-limited quality factor $Q_1$ of 200,000, it exceeds the record-high THG performance reported to date [22] by two orders of magnitude. Unlike microrings and similar waveguide-based devices, where high-power pumping is significantly limited by the coupling into the waveguide, our PhC cavity can be excited directly from free space, enabling much higher-power pumping and improved efficiency. The design is scalable via the lattice constant $a$, and the broad transparency of diamond allows a single geometry to be mapped onto fabricable wavelengths for a range of applications, from telecom-to-visible conversion to color-center-resonant and deep-UV outputs.

Table 1. Comparison of THG performance with related works.

| [Ref] Material / structure | Pump intensity (MW/cm$^2$) | Conversion efficiency $\eta_{conv} = P_{THG}/P_{pump}$ | Device performance factor $\eta_{conv} / (P_{pump}\cdot(\chi^{(3)})^2)$ |
|---|---|---|---|
| [14] Si metasurface (sim.) | 30 | $3.73 \times 10^{-3}$ | $2.07 \times 10^{33}$ |
| [15] Si metasurface (sim.) | 1 | $5.3 \times 10^{-2}$ | $8.83 \times 10^{35}$ |
| [16] Si metasurface (exp.) | 5,850 | $1.0 \times 10^{-5}$ | $2.85 \times 10^{28}$ |
| [19] Si nanodisk (exp.) | 3,000 | $4.0 \times 10^{-5}$ | $2.22 \times 10^{29}$ |
| [17] Ge nanodisk (exp.) | 15,000 | $1.0 \times 10^{-6}$ | $1.85 \times 10^{22}$ |
| [22] AlN microring (exp.) | 17.4 | $1.0 \times 10^{-3}$ | $4.08 \times 10^{37}$ |
| [13] Bulk diamond (exp.) | 2,550,000 | $7.0 \times 10^{-3}$ | $6.22 \times 10^{32}$ |
| (This work, $Q_1$ = 4000) Diamond PhC cavity | 2.2 | $2.7 \times 10^{-7}$ | $2.78 \times 10^{34}$ |
| (This work, $Q_1$ = 200,000) Diamond PhC cavity | 2.2 | $3.4 \times 10^{-2}$ | $3.50 \times 10^{39}$ |

## 6. Conclusion

In conclusion, we have proposed and numerically demonstrated doubly resonant third-harmonic generation in a centrosymmetric diamond photonic crystal slab. Combining a K-point band-edge mode at the fundamental frequency with a Γ-point BIC at the third harmonic, and deforming the circular holes into equilateral triangles that break the lattice symmetry, we simultaneously activate a symmetry-forbidden nonlinear overlap for a realistic slab thickness. A closed-form expression for the coupled-mode-theory THG efficiency provided a material-independent figure of merit that guided both the unit-cell and the heterostructure-cavity design. Three-dimensional FDTD simulations of the designed cavity, which co-confines the fundamental and third-harmonic modes, yield $\eta = 2.7 \times 10^{-7}$ W$^{-2}$, projected to reach ~0.034 W$^{-2}$ at the fabrication-limited quality factor. Because the scheme relies only on lattice symmetry and a material-independent figure of merit, it offers a general and scalable route to monolithic frequency conversion, which is readily transferable to other wavelengths, $\chi^{(3)}$ processes, and material platforms for nonlinear-optical and color-center quantum applications in diamond.

**CRediT authorship contribution statement**

**Sangmin Ji:** Conceptualization, Methodology, Software, Investigation, Formal analysis, Visualization, Writing – original draft. **Satoshi Iwamoto:** Supervision, Writing – review & editing. **Hideki Gotoh:** Supervision, Project administration, Writing – review & editing.

**Acknowledgements.**

This work was supported by Iketani Science and Technology Foundation (0361158-A); Hirose Foundation (12th Research Grant); JSPS KAKENHI (23K19197, 24K17579).

**Disclosures.**

The authors declare no conflicts of interest.

**Data availability.**

Data underlying the results may be obtained from the authors upon reasonable request.

**Declaration of generative AI and AI-assisted technologies in the manuscript preparation process.**

During the preparation of this work the authors used Claude AI and Gemini AI in order to review the grammar and the correspondence between figures and main text. After using this tool/service, the authors reviewed and edited the content as needed and take full responsibility for the content of the published article.